# Lighting Control using Pressure-Sensitive Touchpads

**Alexander Haubold**
Department of Computer Science, Columbia University
500 W. 120th St., 450 CS Building
New York, NY 10027 USA
+1 (212) 939 7146
ahaubold@cs.columbia.edu

**ABSTRACT**
We introduce a novel approach to control physical lighting parameters by means of a pressure-sensitive touchpad. The two-dimensional area of the touchpad is subdivided into 5 virtual sliders, each controlling the intensity of a color (red, green, blue, yellow, and white). The physical interaction methodology is modeled directly after ubiquitous mechanical sliders and dimmers which tend to be used for intensity/volume control. Our abstraction to a pressure-sensitive touchpad provides advantages and introduces additional benefits over such existing devices.

**Keywords**
touchpad, lighting control, intensity control, virtual slider, virtual dimmer

**INTRODUCTION**

Mechanical sliders and dimmers are used in a variety of applications, most predominantly for light dimming controls and in sound mixers. Their use is well-adopted in the physical world and the virtual world of graphical user interfaces (GUIs). However, physical sliders suffer from characteristics typical to mechanical devices with moving parts: frequent usage and age impact the effectiveness and reliability. Their virtual counterparts found in GUIs do not share the same problem and present additional features, such as setting discrete values instantaneously. We have therefore implemented non-mechanical physical sliders and we show their use in the specific domain of lighting control.

In related work, pressure-sensitive touchpads have been investigated for domains outside of their original intended use as pointing devices. Using a much larger touchpad, the SpeechSkimmer [1] implements several vertical regions to allow for real-time control of speed and detail of audio. In [2] a touchpad is divided into 4 vertical strips, where each strip can act as a different GUI widget, including buttons, spinning wheels, and sliders. Finally, remote control devices have been augmented with touchpads in [3] for a new kind of input method that takes advantage of unistrokes.

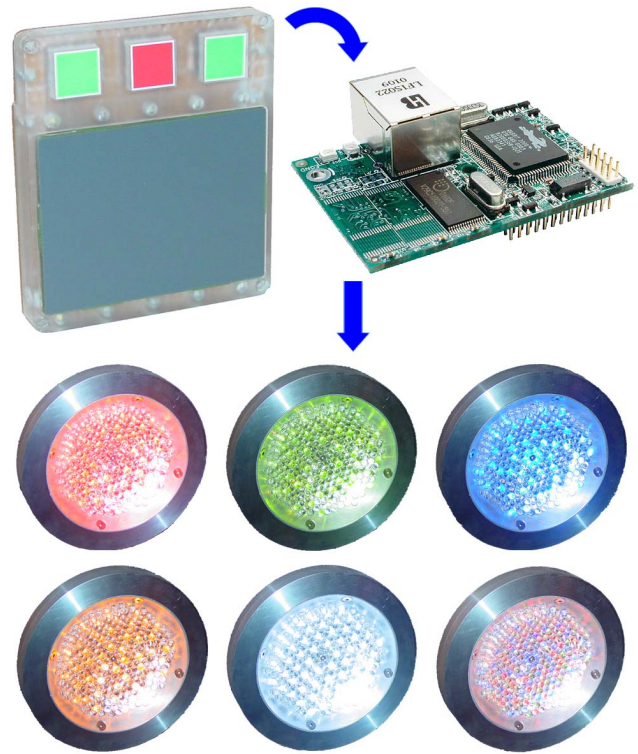

Figure 1: Using the touchpad embedded in a plastic casing (left upper corner), the intensity of each color in a multi-colored LED light cluster (red, green, blue, yellow, and white in one device) can be controlled individually. The interface between touchpad and LED cluster is established using the RabbitCore RCM2200 microprocessor module (right upper corner).

**DESIGN**

Our design includes a touchpad as an input device, and a multi-colored LED light cluster as an output device. For our purposes, the touchpad is divided into 5 vertical regions (sliders), each of which represents a separate physical dimmer (Figure 2). The width of each slider approximates the width of the effective contact area of a finger used to control a mechanical dimmer or a pointing device. Using the 5 sliders, we are able to control the intensity of 5 colors in the LED cluster: red, green, blue, yellow, and white (Figure 1). While red, green, and blue are theoretically sufficient to produce any other color in the spectrum of visible light, we have found that separately adding yellow and white improves the blending of colors on the scale of our LED cluster. Finer-grained spectra of colors beyond the 3 colors of red, green, and blue have already been used in domains such as color matching [4].

Our prototype is implemented with a standard pressure-sensitive touchpad found in most notebook computers as a pointing device. Specifically, we are using a Synaptics model TM1202SBU200-1 measuring 65 mm in width and 49 mm in height. The serial protocol touchpad is connected to a RabbitCore RCM2200 microprocessor module [7]. The custom-designed program on this C-programmable device converts absolute coordinate values from the touchpad into commands understood by the light cluster. The raw horizontal range from 0 to 6143 is reduced to 5 horizontal positions including 4 spatial gaps, and the raw vertical range from 0 to 6143 is mapped to 23 discrete values: 0 to 22 for light intensity.

### INTERACTION

The intensity of a color is changed by sliding the index finger vertically over the touchpad. Intensity of the color increases towards the top and decreases towards the bottom of the sliding region. Similar to mechanical dimmers and GUI-style sliders, the position on the slider is measured absolutely. Using this methodology, discrete intensity levels can be set quickly by simply touching the slider at the desired vertical position.

Since there are no physical boundaries between sliders for different colors, but instead virtual gaps, multiple sliders can be operated at roughly the same time with minimal additional effort per slider. In this scenario, the finger can be moved unobtrusively between the sliders in a horizontal manner to control several color intensities at the same time.

While there exists no indicator on the touchpad as to what values have been set, immediate feedback is present in the form of color brightness by the light cluster. With a processing rate of up to 40 touchpad events per second [5] and a cumulative transmission speed of approximately 40 milliseconds per event from touchpad to the light cluster, there exists virtually no lag between the input and the output devices.

We have asked colleagues to try out the touchpad for the specific task of lighting control. After a 10 second explanation, usage of the device became an effortless and appealing exercise. However, everyone agreed that markings on the touchpad would be a valuable addition.

### CONCLUSION AND FUTURE WORK

We have demonstrated how a standard touchpad unit can be used for lighting control by superimposing 5 virtual vertical sliders onto its surface. The device was implemented without dependence on a full computer system and works as a stand-alone module.

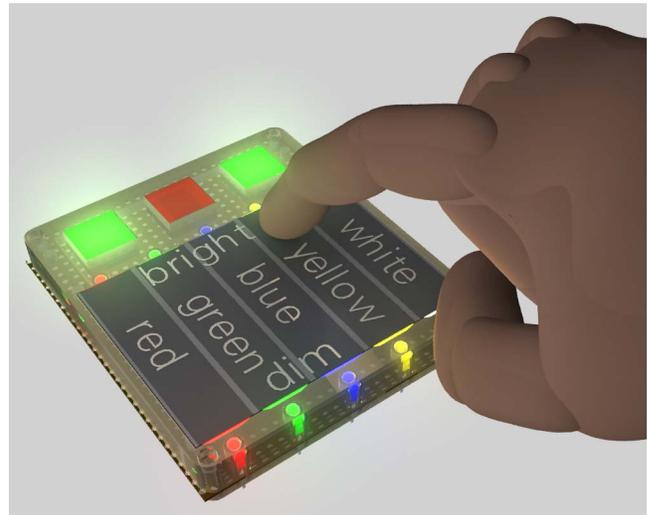

Figure 2: The touchpad is virtually divided into 5 vertical sliders, each of which controls one color. To increase the intensity of a color, the finger is moved towards the top of a slider.

In future work, we plan on incorporating new developments in touchpad technology to provide more visual feedback immediately on the device. Using the transparent touchpad with integrated LCD display [6], slider values can be displayed below the touchpad surface to create a widget similar to the GUI-style counterpart.

### ACKNOWLEDGMENTS

We would like to thank Jeffrey Malins for generously providing technical assistance and his expertise in electronic design, in particular by programming the LED light clusters with PIC17C44-33 microcontrollers.